\documentclass[prl,twocolumn,aps,psfig]{revtex4}

\usepackage{graphicx,color,epsfig,rotate}

\begin{document}
%\draft

\title{On the electronic structure of electron doped LaOFeAs as seen by X-ray absorption spectroscopy}

\author{T. Kroll$^{*,1}$, S. Bonhommeau$^{2}$, T. Kachel$^{2}$, H.A. D\"urr$^{2}$,
J. Werner$^{1}$, G. Behr$^{1}$, A.Koitzsch$^{1}$, R.
H\"ubel$^{1}$, S. Leger$^{1}$, R. Sch\"onfelder$^{1}$, A.
Ariffin$^{3}$, R. Manzke$^{3}$, F.M.F. de Groot$^{4}$, J.
Fink$^{1,2}$, H. Eschrig$^{1}$, B. B{\"u}chner$^{1}$, and M.
Knupfer$^{1}$}

\affiliation{$^{1}$ IFW Dresden, P.O. Box 270116, D-01171 Dresden,
Germany} \affiliation{$^{2}$ Berliner
Elektronenspeicherring-Gesellschaft f\"ur Synchrotronstrahlung
m.b.H. (BESSY), Albert-Einstein-Strasse 15, 12489 Berlin, Germany}
\affiliation{$^{3}$Humboldt-Universit\"at zu Berlin, Institut
f\"ur Physik, Newtonstrasse 15, 12489 Berlin, Germany}
\affiliation{$^{4}$ Department of Inorganic Chemistry and
Catalysis, Utrecht University, Sorbonnelaan 16, 3584 CA Utrecht,
Netherlands}

\date{\today}
\begin{abstract}
We investigated the recently found superconductor $\rm
LaO_{1-x}F_xFeAs$ by X-ray absorption spectroscopy (XAS). From a
comparison of the O $K$--edge with LDA calculations we find good
agreement and are able to explain the structure and changes of the
spectra with electron doping. An important result from this edge
is a limitation of the Hubbard $U$ to values not significantly
larger than 1\,eV. From experimental Fe $L_{2,3}$--edge spectra
and charge transfer multiplet calculations we gain further
information on important physical values such as hopping
parameters, the charge transfer energy $\Delta$, and the on-site
Hubbard $U$. Furthermore we find the system to be very covalent
with a large amount of ligand holes. A shift in the chemical
potential is visible in the O $K$- and Fe $L_{2,3}$--edge spectra
which emphasizes the importance of band effects in these
compounds.
\end{abstract}

 \pacs{etwas}

\maketitle

%\section{Introduction}

Recent reports of superconductivity in $\rm LaO_{1-x}F_xFeAs$
\cite{kamihara_JACS08,takahashi_nature08} stimulated enormous
scientific efforts. In fact, a large oxypnictides family $\rm
Ln(O_{1-x}F_x)FeAs$ (Ln=La, Ce, Pr, Nd, Sm, Gd) has been found to
be superconducting with a transition temperature up to 55\,K
\cite{Wen_EPL08, Chen_CM08, Chen_CM08b, Ren_CM08, Liu_CM08} and
high upper critical fields \cite{Hunte_CM08}. These compounds are
particularly interesting as they are the first example of possible
unconventional superconductivity with a large transition
temperature in a non-cuprate transition metal compound. Their
crystal structure is similar to high-$T_c$ cuprates. Both have a
layered structure in which the role of the $\rm CuO_2$ planes is
now played by FeAs, mediated by (doped) LnO layers that behave as
charge reservoirs.

\par

In order to gain a better understanding of these new compounds,
various experiments and theoretical calculations have been carried
out. Possible unconventional multiband behavior
\cite{Cao_CM08,Eschrig_CM08} and evidence for gap nodes
%\cite{shan,mu,ren_e} have been considered from experiments and
\cite{shan,mu} have been considered from experiments and
theoretical studies, while various scenarios for the
superconducting mechanism have been discussed
%\cite{Boeri_CM08,Mazin_CM08,Lee_CM08,Han_CM08,Dai_CM08,Kuroki_CM08}.
\cite{Boeri_CM08,Mazin_CM08,Lee_CM08,Kuroki_CM08}. Nuclear
Magnetic Resonance (NMR) measurements find evidence for line nodes
in the superconducting gap and spin--singlet pairing in the
superconducting state \cite{Grafe_CM08} while Andreev spectroscopy
finds a single nodeless BCS like gap \cite{Chen_CM08c}. Electronic
structure calculations within the local--density--approximation
(LDA) show a high density of Fe 3$d$ states near the Fermi level,
whereas the density of states for all other ions is low
\cite{Singh_CM08, Boeri_CM08,Cao_CM08,Kuroki_CM08,Eschrig_CM08}.
%\cite{Singh_CM08, Xu_CM08,
%Boeri_CM08,Cao_CM08,Ma_CM08,Kuroki_CM08,Eschrig_CM08}
LDA plus
dynamical--mean--field--theory (DMFT) as well as LDA calculations
that include Hubbard's $U$ conclude an intermediate strength of
$U\sim 1-4$\,eV and Hund's exchange coupling $J_H\sim 0.3-0.9$\,eV
at the Fe site
\cite{Cao_CM08,Haule_CM08,Shorikov_CM08,Si_CM08,Daghofer_CM08,Haule_CM08b,Kurmaev_CM08},
 and the undoped system has been predicted to exhibit a bad metallic behavior
\cite{Haule_CM08,Craco_CM08}.

\par

Core level spectroscopic measurements such as X--ray absorption
spectroscopy (XAS) are appropriate experimental methods to shed
new light on this topic. In XAS, a core electron is excited into
an unoccupied state near the Fermi level, i.e. one probes empty
states. Combined with a theoretical description that takes the
core hole and other contributions properly into account, it can
provide valuable information on the electronic structure of the
investigated system, such as the Hubbard $U$, the charge transfer
energy $\Delta$, Hund's coupling $J$, or the electronic hopping
parameters. In this article, we present experimental data from O
$K$- and Fe $L_{2,3}$--absorption edges. For these measurements we
chose undoped LaOFeAs and electron doped $\rm LaO_{1-x}F_xFeAs$
polycrystalline samples in a doping range between $x=0.0$ and
$0.15$. These data will be discussed and compared to LDA
calculations as well as to Fe 3$d$ charge transfer multiplet
calculations.

\par

%\section{Experimental methods}

Polycrystalline samples of $\rm LaO_{1-x}F_xFeAs$ were prepared as
pellets as described in Ref. \onlinecite{Drechsler_CM08}, and
consist of 1 to 100\,$\mu$m sized grains of $\rm
LaO_{1-x}F_xFeAs$. Their X-ray absorption spectroscopy studies at
the O $K$- and Fe $L_{2,3}$--edge were performed at the PM3
beamline of the synchrotron radiation source Bessy II. The energy
resolution was set to 180 and 300\,meV at 530 and 710\,eV photon
energy, respectively. Data were recorded by measuring the
fluorescence signal. Compared to electron yield data, the
penetration depth is larger for fluorescence which is more
preferable in this case. In order to obtain appropriate surfaces,
the pellets were filed {\it in-situ} in a vacuum environment of
$1\times 10^{-7}$\,mbar and measured at $1\times 10^{-8}$\,mbar.

\par

%\section{Results and discussion}

%\subsection{Oxygen $K$-edge}

\begin{figure}[t]
\begin{center}
\includegraphics[width=0.85\columnwidth, angle=0, clip]{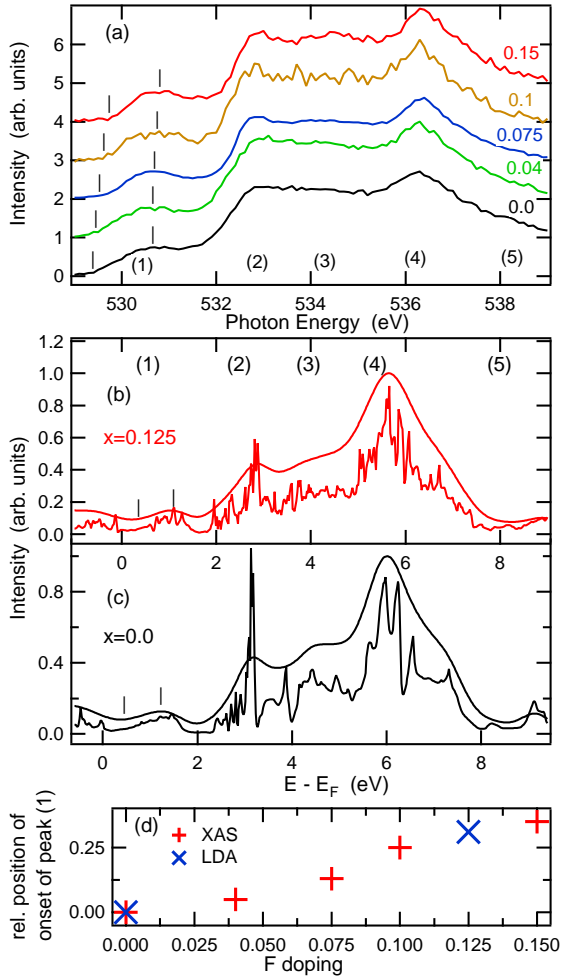}
\end{center}
\caption{\small (Color online) $\rm LaO_{1-x}F_xFeAs$: (a) Doping
dependence of XAS O $K$--edge spectra. The spectra have been
normalized at 610\,eV where they become structure less and doping
independent. (b) and (c) Oxygen $p$--projected partial density of
states for $x=0.0$ and $x=0.125$. (d) Energy shift of the onset of
peak (1) as compared to the undoped sample for experimental and
theoretical results.} \label{O_K_data}
\end{figure}

In Fig. \ref{O_K_data}(a) O $K$--edge spectra are shown for photon
energies between 529 and 539\,eV. This region can be assigned to
excitations from the O 1$s$ core level into unoccupied O 2$p$
states. In the XAS spectra the onset of peak (1) shifts by
0.35\,eV towards higher photon energies with doping
\cite{footnote1}, whereas peak (1) itself shifts only by
$\approx$100\,meV and peak (2) (and also peak (4)) do not shift.
From X-ray photoemission spectroscopy (XPS) experiments
\cite{Koitzsch_CM08} it has been observed that the La 4$d$ level
shifts relative to the chemical potential by about 200\,meV from
$x=0.0$ to $x=0.1$ while the As 3$d$ level hardly shifts. This can
be ascribed to a change of the Madelung potential between the As
and La layers upon doping.

\par

Further insight is gained from LDA calculations
\cite{Eschrig_CM08} as shown in Fig. \ref{O_K_data}(b) and (c) for
different doping level.
%High precision density
%functional calculations were performed with the local density
%approximation functional of [12] using the full-potential
%local-orbital code [13] in the version FPLO7-28 [14] with its
%default orbital settings. The effect of fluor doping for oxygen is
%studied by replacing one of eight oxygen atoms with a fluor atom
%in an ordered way.
%described in more detail in Ref. \onlinecite{Eschrig_CM08} (see
%Fig. \ref{O_K_data}).
The oxygen $p$--projected density of states (PDOS) has been
broadened by a Gaussian with FWHM of 0.18\,eV and a Lorentzian
with FWHM of 0.2\,eV. From the LDA PDOS, peak (1) can be ascribed
to a hybridization of O states with Fe 3$d$ states and peak (2) to
hybridization with La 4$f$ states. The position of peak (1)
essentially measures the Madelung potential at the Fe layers,
while the position of peak (2) is related to the Madelung
potential at the La layer. One effect of doping is a change in the
Madelung potentials. According to LDA, the Madelung potential for
La sites should hardly change as compared to the O site, whereas
for Fe it leads to a shift of peak (1) towards higher energies,
which is corroborated by the XAS data at the O $K$--edge. A second
and more important effect with electron doping is a shift of the
chemical potential due to a change in the carrier number. By this,
the onset of peak (1) is further shifted to higher energies as
compared to the La peak.

\par

When comparing the experimental and theoretical results, one
observes first of all that the overall agreement is good , which
tells that apparently the core hole effect in the absorption
process is small. Therefore  a direct interpretation of XAS
measurements in terms of the partial unoccupied density of states
is possible, analogous to a one electron addition process
\cite{deGroot_CMR05}. Note that the energies as given in Fig.
\ref{O_K_data}(a) and (b,c) correspond to different reference
values. In XAS the photon energy relates to the energy difference
between the core level and the unoccupied states, whereas in the
DOS the energy difference between the Fermi level and the
unoccupied states is given. For the sake of clarity the energy
regions in Fig. \ref{O_K_data}(b) and (c) are chosen in such a way
that peak (2) aligns with the experimental data. When focussing on
the onset of peak (1), i.e. on the change in the chemical
potential, a clear doping dependence is observed. This is
illustrated in Fig \ref{O_K_data}(d) where the shift of the onset
of peak (1) as compared to $x=0.0$ is shown. Such an increase is
supported by the PDOS since the shift in the onset of peak (1)
\cite{footnote1} between $x=0.0$ and $x=0.125$ matches well to the
slope found from the experimental data. This agreement between
theory and experiment stresses the observation that the
experimental O $K$--edge is strongly affected by the shift of the
chemical potential with doping.

\par

Furthermore, in LDA no on-site Hubbard $U$ is taken into account.
When switching it on at the Fe site, this will have an effect on
the energetic position of the Fe 3$d$ spectral weight. As the
relative position of the O $K$ XAS peaks matches those determined
by LDA calculations within one eV, the on-site Hubbard $U$ can be
expected not to be significantly larger than 1\,eV. Note that the
strong electronic correlations in cuprates such as $\rm
La_{2-x}Sr_xCuO_2$, are clearly visible in the O $K$--absorption
edges \cite{fink1994}.

%\subsection{Iron $\bf L_{2,3}$-edge}

In order to gain more information on the electronic structure of
this system, XAS experiments at the Fe $L_{2,3}$--edge have been
performed. According to the dipole selection rules, the Fe
$L_{2,3}$ absorption edges correspond to excitations of Fe 2$p$
core level electrons into unoccupied Fe 3{\it d} electronic
states. Upon variation of the energy of the incident light,
different Fe 3{\it d} orbitals can be probed \cite{deGroot_PRB90}.
Contrary to what occurs at the O $K$--edge, core holes cannot be
neglected for Fe 3$d$ excitations. Therefore, simulations of the
Fe $L_{2,3}$--edge require consideration of multiplet splitting,
hybridization, and crystal field effects. Fe $L_{2,3}$--edge
experimental data have been taken by recording the fluorescence
yield and corrected for a self--absorption process in the
fluorescence signal \cite{Troeger_PRB92}. Self--absorption is
stronger at the $L_3$ edge, which cannot be fully corrected and
thereby the $L_3$--edge intensity will be somewhat underestimated
as compared to the $L_2$--edge. All data have been normalized at
750\,eV where the signal corresponds to excitations into continuum
states.

\par

\begin{figure}[t]
\begin{center}
\includegraphics[width=0.85\columnwidth, angle=0, clip]{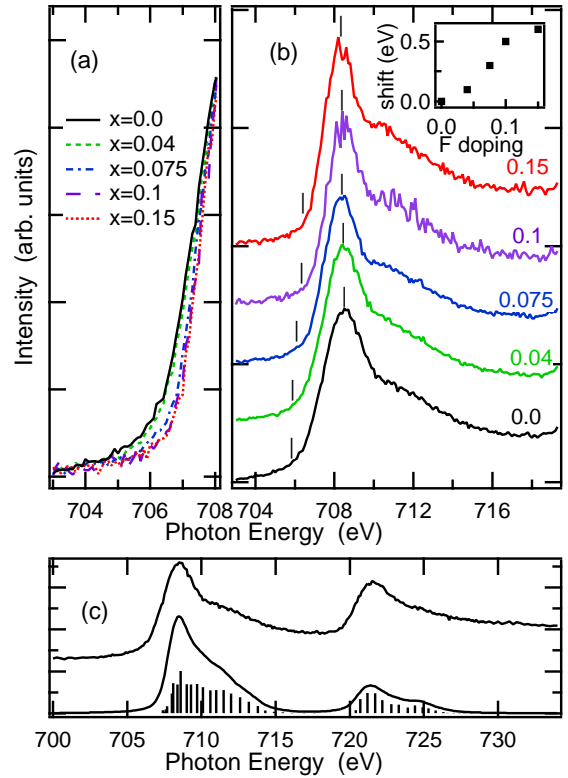}
\end{center}
\caption{\small (Color online) $\rm LaO_{1-x}F_xFeAs$: Doping
dependence of the Fe $L_{2,3}$--edge. (a) Blowup of the low energy
side of experimental Fe $L_3$--edge. Note that for the sake of
clarity, the spectra in Fig. (a) were shifted in such a way that
their maxima match. (b) Experimental $L_3$--edge for the same
doping range. The inset shows the shift of the onset of the main
peak in eV relative to $x=0.0$ as a function of doping $x$. (c)
Comparison of LaOFeAs XAS spectra with charge transfer multiplet
calculation, for values see text.} \label{Fe_L}
\end{figure}

In Fig. \ref{Fe_L}(a) and (b) the experimental Fe $L_3$--edge XAS
spectra for different doping levels are shown. Two main changes
appear with F doping. The energy position of the main peak around
708\,eV shifts slightly with doping towards lower energies. This
shift amounts to $\approx$150\,meV on going from $x=0.0$ to
$x=0.15$ and can be explained by the observation that the XPS Fe
2$p$ core level excitations do not shift relative to the chemical
potential with doping within the experimental resolution
\cite{Koitzsch_note}, while the chemical potential shifts by
$\approx$200\,meV with doping from x=0.0 and
x=0.2\cite{Koitzsch_CM08}. In other words, this excitation energy
as seen in Fig. \ref{Fe_L} decreases upon doping. Moreover, the
onset of the $L_3$--edge \cite{footnote1} shifts to higher photon
energies by $\approx$600\,meV, similar to the O $K$--edge. Such a
shift could also cause an asymmetric peak narrowing and affect the
position of the peak maxima. In Fig. \ref{Fe_L}(a) a blowup of the
low energy side of the $L_3$--peak is shown. Note that for the
sake of clarity, the spectra only in Fig. \ref{Fe_L}(a) were
shifted to the same energy of their maxima. A shifted onset is
consistent with additional electrons at the Fe sites which
diminishes the number of holes, i.e. the total intensity at the Fe
$L$--edge. Therefore, the doped electrons reside (partially) on
the Fe sites. These findings are supported by valence band
photoemission spectroscopy. When doping with electrons, an
increase in intensity just below the Fermi energy where the Fe
states are located has been observed \cite{Koitzsch_CM08}. Since
the conduction band is partially filled, this shift leads to the
experimentally observed effect on the Fe $L_{2,3}$--edge and
emphasizes the importance of band effects on the absorption edge
in these compounds. Similar to what has been observed at the O
$K$--edge, the onset of the spectra shifts monotonically to higher
energies (see inset in Fig. \ref{Fe_L}(b)) and stresses the shift
of the chemical potential as the origin of both effects.

%When comparing these spectra to literature, one finds very similar
%5spectra for arsenoprite FeAsS \cite{Mikhlin_PCM05}. Here, the Fe
%ion is surrounded by six ligands in a cubic environment. In
%general, the spin state of FeAsS has been accepted as low spin S=0

Although As is formally $\rm As^{3-}$, its 4$p$ shell is not
completely filled due to hybridization with the Fe 3$d$ shell. The
radial function of 4$p$ states is far ranging compared to the 2$p$
states of O as in copper oxides. Therefore, a large overlap of the
wave functions between Fe and As occurs which leads to a strong
hybridization and a further delocalization of the Fe 3$d$
electrons. With F doping delocalized additional charge is
introduced into this layer. When looking at the crystal structure,
the FeAs layers consist of Fe ions on a square lattice that are
tetrahedrally surrounded by four As ions. These As tetrahedra are
slightly tetragonally distorted which reduces the local symmetry
from $T_d$ to $S_4$.
%In this lower symmetry, the formerly degenerate $d_{xy}$, $d_{yz}$,
%$d_{zx}$ orbitals split into a single $d_{xy}$ and a two--fold
%degenerate $d_{yz}$, $d_{zx}$ levels. Similarly the degenerate
%$d_{x^2-y^2}$ and $d_{3z^2-r^2}$ orbitals (in $T_d$ symmetry)
%split. Looking at $T_d$ symmetry, the symmetry elements in $O_h$
%and $T_d$ are identical beside the fact, that the sixfold $C_4$
%and $C_2$ rotations in $T_d$ symmetry contain an inversion center.
%This leads to three main changes: Due to the inversion center, the
%gerade/ungerade differentiation is lost (i.e. $T_{2g}\rightarrow
%T_2$ and $E_g\rightarrow E$), the energy level $E$ and $T_2$ are
%inverted (i.e. negative crystal field), and finally the hopping
%relation change as described above.
%In total, simulations in $O_h$ symmetry
%can be safely used for $T_d$ symmetry when taking their
%differences carefully into account.
The crystal field splitting $10Dq$ has been predicted to be rather
small ($\sim 0.25$\,eV) \cite{Shorikov_CM08,Cao_CM08,Haule_CM08b}.
From LDA plus DMFT calculations the $d_{x^2-y^2}$ and
$d_{3z^2-r^2}$ orbitals have been found to be still degenerate
\cite{Craco_CM08}, and the $E_g$/$T_{2g}$ splitting to be dominant
over the tetragonal distortion. Therefore, it is reasonable to use
$T_d$ local symmetry as a starting point for simulations of the Fe
$L$--edge. In an X-ray absorption excited state, a 2$p$ core hole
has been created which interacts with valence electrons via
Coulomb interactions. Its energy scale is in the range of a few eV
and leads to an additional multiplet splitting
\cite{Zaanen_PRB85}. This interaction justifies a local
description also of a metallic system as long as it is strong
enough to lift the excited states out of the conduction band.

\par

In order to interpret the experimental data further, we performed
charge transfer multiplet calculations for divalent $\rm Fe^{2+}$
(3$d^6$) \cite{deGroot_PRB90,Butler,Cowan,Thole_PRB85}. The
calculations have been done for $T_d$ symmetry and the hopping
relation $V_{e}=\frac{1}{\sqrt 3} V_{t_2}$, assuming the relation
$pd\sigma=-\frac{\sqrt 3}{4}pd\pi$. Note that a band effect such
as the shift of the chemical potential is beyond this local
approach. In Fig. \ref{Fe_L}(c) a comparison of the XAS
fluorescence data and charge transfer multiplet calculations is
shown. The agreement between experiment and theory at the
$L_{2,3}$--edge is good. Note that the intensity of the
$L_2$--edge is overestimated by the self--absorption correction.
The parameter set that reproduces the experimental data best is
$10Dq=0.2$\,eV, $\Delta=d^{7}\underline{L}-d^6=1.25$\,eV
($\underline{L}$ denotes a ligand hole), $U=1.5$\,eV, and
$|pd\pi|=0.27$\,eV. The core hole potential $Q$ is normally about
$1-2$\,eV larger than $U$ and has been set to $Q=U+1$\,eV. The
Slater-Condon parameters have been reduced to 80\% of their
Hartree--Fock values as it is reasonable in solids, which leads to
the two Hund's couplings $J_{e_g}=0.90$\,eV and
$J_{t_{2g}}=0.78$\,eV for the ground state. The shoulder at
$\approx712$\,eV is provoked by charge transfer effects and
emphasizes the hopping values above. A square band containing 5
states and a bandwidth of 2.5\,eV for the ligand hole state has
been added in the charge transfer calculations. The multiplet
intensities have been broadened by a Gaussian (0.3\,eV) and a
Lorentzian (0.6\,eV at $L_3$ and 0.8\,eV at $L_2$ due to different
lifetime broadenings). Since the core hole potential is rather
small ($\approx$ 2.5\,eV), the excited states are not shifted far
out of the Fe 3$d$ band at $\approx 2$\,eV above the Fermi energy,
and therefore band effects become visible.

When writing the wave function as a sum of three configurations
with different ligand holes $\psi=a d^6 + b d^{7}\underline{L}^1 +
c d^{8}\underline{L}^2$ ($a^2+b^2+c^2=1$), the hole occupation can
be given. Using the parameters as written above, it follows that
$a^2$=0.558, $b^2$=0.393, and $c^2$=0.049. For all possible
parameter sets, a high spin situation $S=2$ has been determined in
agreement with LDA+DMFT \cite{Craco_CM08}. The energy difference
to the intermediate spin state $S=1$ is $\approx 0.4$\,eV which is
enough to omit the role of this latter state. This can be
understood by the small values of the crystal field and hopping
parameters. A low ligand field splitting involves the filling of
energy levels following Hund's rule and forces the system into a
high spin configuration.

\par

%Upon doping the system with electrons, the lattice parameters $a$
%and $c$ shrink \cite{Liu_CM08,Hamann_CM08}, which leads to an
%increase in hybridization and a larger crystal field. This changes
%are in the order of 1\%, which makes them invisible in this
%experiment since $t_{pd}\sim r^{-7/2}$ and $10Dq\sim r^{-1}$. Note
%that a larger crystal field splitting shifts the multiplet spectra
%slightly to lower energies, consistent with the experimental
%observation.

%Since the core hole potential is rather small ($\approx$ 2.5\,eV),
%the excited states are not shifted far below the Fe 3$d$ band at
%$\approx 2$\,eV above the Fermi energy, and therefore band affects
%are taken into account.

%\section{Summary}

In summary, from X-ray absorption spectroscopy measurements
together with charge transfer multiplet and LDA calculations,
deeper insight into the electronic structure of $\rm
LaO_{1-x}F_xFeAs$ has been proposed. The O $K$--edge is well
described by LDA calculations. The influence of the Madelung
potential on different ions coincides in both experimental and
theoretical spectra. The shift in the chemical potential is
clearly visible in the absorption edge. Furthermore, an upper
limit of the on-site Hubbard $U$ could be assessed to $\approx
1$\,eV.

\par

Band effects have a significant influence also on the shape of the
experimental Fe $L$--edge absorption spectra. A shift in the
chemical potential towards higher energies is observed in
agreement with the results of the O $K$--edge, which stresses the
existence of a low Hubbard $U\approx 1$\,eV. Further valuable
information could be extracted from Fe $L_{2,3}$--edge absorption
spectroscopy together with charge transfer multiplet calculations
in $T_d$ symmetry. The extracted parameter set appears to be
similar to what has been suggested from DFT calculations,
especially the low crystal field splitting \cite{Shorikov_CM08}
and the small hopping parameters \cite{Daghofer_CM08}. The low
Hubbard $U$ fits to the upper bound as concluded from the
comparison between O $K$-edge XAS spectra and DOS. Furthermore,
due to small values of the charge transfer energy $\Delta$ and the
Hubbard $U$ the system turns out to be very covalent similar to
what has been predicted from DFT calculations
\cite{Singh_CM08,Haule_CM08}.

%\section*{Acknowledgment}
This investigation was supported by the DFG (SFB 463 and KR
3611/1-1) and DFG priority program SPP1133.


\begin{thebibliography}{00}
\bibitem{footnote1}The onset of the peaks has been defined as the
photon energy at which the intensity reaches 10\% of its value at
the peak maximum
\end{thebibliography}
\end{document}